# Multilink and AUV-Assisted Energy-Efficient Underwater Emergency Communications

Zhengrui Huang, *Student Member*, *IEEE*, and Shujie Wang

*Abstract*—Recent development in wireless communications has provided many reliable solutions to emergency response issues, especially in scenarios with dysfunctional or congested base stations. Prior studies on underwater emergency communications, however, remain under-studied, which poses a need for combining the merits of different underwater communication links (UCLs) and the manipulability of unmanned vehicles. To realize energy-efficient underwater emergency communications, we develop a novel underwater emergency communication network (UECN) assisted by multiple links, including underwater light, acoustic, and radio frequency links, and autonomous underwater vehicles (AUVs) for collecting and transmitting underwater emergency data. First, we determine the optimal emergency response mode for an underwater sensor node (USN) using greedy search and reinforcement learning (RL), so that isolated USNs (I-USNs) can be identified. Second, according to the distribution of I-USNs, we dispatch AUVs to assist I-USNs in data transmission, i.e., jointly optimizing the locations and controls of AUVs to minimize the time for data collection and underwater movement. Finally, an adaptive clustering-based multi-objective evolutionary algorithm is proposed to jointly optimize the number of AUVs and the transmit power of I-USNs, subject to a given set of constraints on transmit power, signal-to-interference-plus-noise ratios (SINRs), outage probabilities, and energy, which achieves the best tradeoff between the maximum emergency response time (ERT) and the total energy consumption (EC). Simulation results indicate that our proposed approach outperforms benchmark schemes in terms of energy efficiency (EE), contributing to underwater emergency communications.

## I. Introduction

RECENT advancements in wireless communications have provided potential solutions to energy-efficient emergency responses and post-disaster rescues [1], including monitoring industrial devices, collecting environmental parameters, and recovering communication networks. Most studies focused on above-ground emergency response, such as deploying wireless sensor networks to provide emergency services and scheduling unmanned aerial vehicles to collect emergency data [2], while the prior works on underwater emergency communications are relatively limited. An underwater emergency communication network (UECN) comprises underwater sensor nodes (USNs) and surface sink nodes (SSNs), where a USN transmits data to an SSN (e.g., ship, buoy, drone) or transfers data to a relay node (RN) [3]. Compared to terrestrial communications, underwater communications mean transmitting data through underwater communication links (UCLs), including underwater light (UL), underwater acoustic (UA), and radio frequency (RF) links, and complex underwater environments become the most influential factor [4]. RF waves are easily affected by (sea) water, causing limited communication ranges. The slow propagation speed of acoustic waves in (sea) water causes very high latency of UA channels. Underwater pressure and temperature and organisms misalign transceivers and enhance path losses of UL channels [5]. To mitigate underwater environmental impacts on channels, unmanned vehicles gain attention. For example, an unmanned surface vehicle (USV) and an autonomous underwater vehicle (AUV) function as a surface gateway and an underwater RN, respectively. Motivated by the above, one feasible solution to establishing an energy-efficient UECN is to comprehensively consider relay detection and selection and AUV deployment, which is the focus of this study.

*A. Related Works*

An individual channel has merits and demerits [6]. As listed in Table I, UL links can support high-data-rate communication within a relatively short communication distance. Elamassie et al. [7] first derived a closed-form expression for modeling the bit error rate (BER) of vertical UL links. To extend the optical communication range, a light-path (LP) routing protocol based on a beamwidth tradeoff was proposed to optimize end-to-end (E2E) data rates [8]. In addition, an optical relay system was proposed to approximate outage probabilities [9], and Shihada et al. [10] adopted an LED or laser to provide wireless access at different distances. In particular, UA communications support data transmission up to 20 km, providing wider communication coverage than UL and RF communications. Huang et al. [11] proposed an adaptive modulation scheme for optimal channel selection, where the frequency of UA links ranged from 900 to 1500 Hz. Following [11], an active queue management policy was proposed to schedule acoustic links [12], but this scheme was only suitable for static scenarios. Considering the quality of service, an energy-effective acoustic network was proposed

Wang was supported in part by the National Aeronautics and Space Administration under Grant 80NSSC22K0384 and in part by the National Science Foundation under Grant 2127329. *(Corresponding author: Shujie Wang.)*
Zhengrui Huang is with the Department of Geography, The Pennsylvania State University, University Park, PA 16802, USA (email: zqh5210@psu.edu).
Shujie Wang is with the Department of Geography, Earth and Environmental Systems Institute (EESI), and Institute for Computational and Data Sciences (ICDS), The Pennsylvania State University, University Park, PA 16802, USA (email: skw5660@psu.edu).

to reduce latency and transmission errors [13]. Compared to UL and UA channels, RF channels provide moderate data rates and have a unique propagation mechanism whose communication distance is lower than 100 m under the water. Che et al. [14] investigated the performance of RF underwater communication, indicating that increased RF frequencies reduced propagation distances. Following [14], a time division multiplexing access (TDMA)-based RF network was built for coastal monitoring purposes [15], where the performance of RF signals at 3 kHz was evaluated. Compared to a two-dimensional (2D) network architecture, a three-dimensional (3D) RF network was built to increase data rates [16].

TABLE I: COMPARISONS OF DIFFERENT CHANNELS

| Parameter | UL channel | UA channel | RF channel |
|---|---|---|---|
| Frequency | $10^{12}$-$10^{15}$ Hz | 10 Hz-100 kHz | MHz ranges |
| Transmit power | In the range of milliwatts to watts | Few watts | Few watts |
| Data rate | Gbps | kbps | Mbps |
| Communication distance | 10-100 m | up to 20 km | up to 10 m |
| Performance parameter | Absorption, turbulence, and organic matter | Temperature, shipping activity, salinity, and pressure | Conductivity and permeability |
| Propagation speed | $2.25 \times 10^8$ m/s | 1500 m/s | $2.25 \times 10^8$ m/s |
| Line-of-Sight (LoS)/None LoS (NLoS) | LoS only | Both | Both |

Given [7]-[15], using an individual channel is not robust, i.e., longer communication distances cause lower data rates, which poses a need for hybrid systems [17], including RF-optical, RF-acoustic, and acoustic-optical systems. In [18], a hybrid visible light communication (VLC)-RF relaying system was proposed to support spatially random terminals. To enhance transmission capacity, Luo et al. [19] proposed an adaptive routing scheme for an RF-acoustic system. Similarly, Han et al. [20] integrated optical and acoustic channels to enhance throughput, and Lin et al. [21] proposed a software-defined networking (SDN) system based on UA-UL links, extending communication coverage.

However, those studies did not account for the distribution of USNs. To address this issue, deploying AUVs can be a feasible scheme. An AUV-assisted acoustic network was established to support mobile data collection [22]. Similarly, Han et al. [23] studied a high-availability model to schedule AUVs, increasing packet delivery ratios, and a heterogeneous underwater network was built using stochastic optimization [24]. To realize optimal power control, a reinforcement learning (RL) model was used to maximize signal-to-noise ratios (SNRs) [25]. Furthermore, it should be noted that energy is the most fundamental limitation of AUVs [26]. Carolis et al. [27] proposed a runtime estimation framework to calculate the energy consumption (EC) of AUVs, and Deutsch et al. [28] compared various energy management schemes, including rule-based and optimization-based models.

Therefore, to realize energy-efficient underwater emergency communications, it is crucial to study how to build a UECN by joint link selection and AUV deployment. So far, the existing works [21], [25], and [29] did not synthetically solve the issue mentioned above. In [21], the authors proposed a hybrid UL-UA network for underwater communication without optimizing SNRs. In [25], an RN identification algorithm was proposed to transmit data to a USV, but this method was only suitable for a static scenario, reducing underwater communication distances. Similarly, Xing et al. [29] separately minimized the EC of data transmission and maximized system SNRs, although these two objective functions should be jointly considered.

B. Contributions

Motivated by the merits of three channels and the flexibility of AUVs, we propose a multi-link and AUV-assisted UECN. Compared to the prior works [18]-[29], this study considers a cooperative UECN whose resources, including transmit power, time slots, and bandwidth, are limited during the process of data propagation. Thus, to rationally utilize the limited resources, all system units (USN and AUV) must cooperate and work in their optimal emergency response modes (ERMs). That is, one USN within the communication range of the USV can forward data to the USV directly, but the other USNs disconnected from the USV must transfer data to an RN or an AUV. Here, our goal is to simultaneously minimize the maximum emergency response time (ERT) and the total EC, which can be formulated as a multi-objective optimization problem (MOP), and our main contributions include optimal mode selection, optimal AUV deployment, and optimal time-energy balancing.

• Optimal mode selection: First, we formulate three ERMs, including E2E transmission (ERM 1), RN transmission (ERM 2), and AUV transmission (ERM 3). Second, the optimal ERM of a USN is determined using a stepwise algorithm, including relay detection and selection. In the first step, a greedy search (GS)-based algorithm is proposed to find out all USNs that can transmit data to the USV directly, and these USNs can function as underwater RNs to transfer data returned from the remaining USNs. In the second step, following the distribution of RNs, an RL-based algorithm is proposed to select the best RN for a USN disconnected from the USV, maximizing the channel capacity (CC) and ensuring the connectivity between a USN and an RN. Accordingly, I-USNs can be identified.

• Optimal AUV deployment: To ensure that the data returned from I-USNs can be received by the USV, the optimal AUV deployment is obtained by jointly optimizing the locations and controls of AUVs. First, a stepwise clustering scheme is used to determine the optimal locations of AUVs, which minimizes the data transmission time of I-USNs. Second, AUV velocities can be optimized by convex optimization, and the optimal solution is given by the method of Lagrange's multipliers. As a result, an AUV arrives at the optimal suspension location to receive data returned from I-USNs, which shortens the time for underwater data transmission and movement.

• Optimal time-energy balancing: From the multi-objective optimization perspective, this study aims to determine the best time-energy tradeoff by maximizing the energy efficiency (EE) of the UECN. To this end, an adaptive clustering-based multi-objective evolutionary algorithm (AC-MEA) is proposed to yield the Pareto front (PF) of the MOP, where the number of AUVs and the transmit power of I-USNs are optimized, subject to a given set of constraints on transmit power, signal-to-interference-plus-noise ratios (SINRs), outage probabilities, and energy. Numerical results indicate that our proposed approach achieves higher EE than benchmark schemes.

The remainder of this study is organized as follows. In Section II, the system model and the problem formulation are introduced. In Section III, the optimal mode selection and AUV deployment are realized, and the formulated MOP is solved by our proposed AC-MEA. Simulation results and analyses are shown in Section IV, and Section V concludes this study. The main notations used in this study are shown in Table II.

TABLE II: Main Notations

| Parameter | Description |
|---|---|
| $\mathbb{I}(\cdot)$ | Integer operator |
| $\mathbb{R}(\cdot)$ | Fraction operator |
| $\mathbb{S}(\cdot)$ | Indicator function |
| $\mathbb{E}(\cdot)$ | Expectation |
| $\mathbb{L}(\cdot)$ | Loss function of a neural network |
| $\mathbb{C}(\cdot)$ | Counting function |
| $\mathbb{N}(\cdot)$ | Min-max normalization |
| $\mathbf{Q}(\cdot)$ | Q-value |
| $\zeta(\cdot)$ | Depth-dependent function |
| $\mathcal{L}(\cdot)$ | Lagrangian function |
| $\|\cdot\|$ | Euclidean norm |
| $\|\cdot\|$ | Cardinality |
| $N$ | Number of USNs |
| $M$ | Number of AUVs |
| $\mathbb{P}_{out}^{i,0}$ | Outage probability between USN $i$ and USV 0 |
| $\mathbb{P}_{out}^{k,i}$ | Outage probability between USN $k$ and RN $i$ |
| $\mathbb{P}_{out}^{l,j}$ | Outage probability between I-USN $l$ and AUV $j$ |
| $L_{i,k}^{UL}$ | UL PL between USN $i$ and USN $k$ |
| $L_{i,k}^{UA}$ | UA PL between USN $i$ and USN $k$ |
| $L_{i,k}^{RF}$ | RF PL between USN $i$ and USN $k$ |

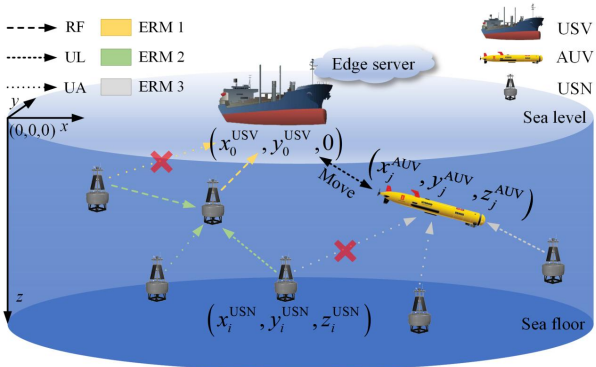

Fig. 1. System model.

## II. System Model and Problem Formulation

The UECN consists of three units, including a set $\mathcal{N}=\{1,...,N\}$ of $N$ USNs, a set $\mathcal{M}=\{1,...,M\}$ of $M$ AUVs, and one USV, as shown in Fig. 1, where USNs are randomly deployed below the sea level to collect data, AUVs are dispatched to assist I-USNs, and the USV works as an SSN, and the locations of USN $i$, AUV $j$, and USV 0 can be denoted by $\left(x_i^{USN}, y_i^{USN}, z_i^{USN}\right)$, $\mathbf{u}_j = \left(x_j^{AUV}, y_j^{AUV}, z_j^{AUV}\right)$, and $\left(x_0^{USV}, y_0^{USV}, 0\right)$, respectively. In our model, we consider the following configurations: 1) The locations of USNs and AUVs are known to the USV, and the USV is responsible for computation-intensive missions [25]; 2) Three types of links are introduced, including UL, UA, and RF links, and the frequency-division multiple access (FDMA) over orthogonal channels is used for underwater communications; 3) The upper bounds of transmit power can be denoted by $P_{max}^{UL}$, $P_{max}^{UA}$, and $P_{max}^{RF}$, respectively, the carrier frequencies are $10^{13}$ Hz, 20 kHz, and 5 MHz [5], respectively, and the bandwidth is 1 kHz; 4) The time domain of a channel is divided into slots, and time slot $t$ is for transmitting a packet. Similar to [14], the duration of $t$, denoted by the ERT of a USN per round, includes transmission time and latency, guard time, and preamble time. Given the predefined configurations, the geographically fixed USNs can transmit their data to the USV if they are within the communication range of the USV; otherwise they must transfer data to RNs or AUVs. Without loss of generality, three ERMs are defined as follows.

1) ERM 1: A USN directly transmits data to the USV through the selected UCL.
2) ERM 2: A USN directly transmits data to the USN in ERM 1 through the selected UCL, and thus the data can be transferred to the USV by a single-hop relay.
3) ERM 3: A USN directly transmits data through the selected UCL to an AUV, and the AUV can move back to the USV to offload data.

### A. Communication Model

To describe the characteristics of different channels, three PL models are selected to reflect the channel attenuation caused by communication distances and frequencies.

The most commonly used channel model of a UL link is the LoS model, where the light beam of a transmitter aligns with the direction of a receiver [8]

$$L_{i,k}^{UL} = 10\log_{10}\left[\frac{\eta_T \eta_R \exp(-c(\lambda)d_{i,k})A_{Rec}}{2\pi d_{i,k}^2 \cos(\theta_{i,k})(1-\cos(\theta_0))}\right], \quad (1)$$

where $d_{i,k}$ is the Euclidean distance between USN $i$ and USN $k$, $\lambda$ is the wavelength, $\theta_{i,k}$ is the elevation angle, $\theta_0$ is the laser beam divergence angle, $A_{Rec}$ is the receiver aperture area, $\eta_T$ and $\eta_R$ are the optical efficiencies of a transmitter and a receiver, respectively, and $c(\lambda)$ is the attenuation coefficient [30]

$$c(\lambda) = a(\lambda) + b(\lambda), \quad (2)$$

where $a(\lambda) = \lim_{\Delta D \to 0} \frac{P_A}{P_I \Delta D}$ and $b(\lambda) = \lim_{\Delta D \to 0} \frac{P_S}{P_I \Delta D}$ denote the coefficients of absorption and scattering, respectively, $\Delta D$ is the water thickness, $P_I$ is the power of incident light, and $P_A$ and $P_S$ are the power of absorption and scattering, respectively.

The attenuation of an acoustic channel can be expressed as [31]

$$L_{i,k}^{\text{UA}} = \kappa 10\log_{10} d_{i,k} + \frac{d_{i,k}}{1000} 10\log_{10} \phi(f), \quad (3)$$

where $\kappa \in [1,2]$ is the spreading coefficient, $f$ is the frequency, and $10\log_{10}\phi(f)$ is given by

$$10\log_{10}\phi(f) = \frac{0.11f^2}{1+f^2} + \frac{44f^2}{4100+f^2} + 2.75 \times 10^{-4} f^2 + 0.003. \quad (4)$$

In addition, the ambient noise of a UA link is an inevitable factor influencing communication performance whose sources include turbulence, shipping, waves, and thermal noise, which can be expressed as a function of $f$ in kHz [13]

$$\begin{cases} 10\log_{10} N_T(f) = 17 - 30\log_{10} f, \\ 10\log_{10} N_S(f) = 40 + 20(s - 0.5) + 26\log_{10} f - 60\log_{10}(f + 0.3), \\ 10\log_{10} N_W(f) = 50 + 7.5\sqrt{w} + 20\log_{10} f - 40\log_{10}(f + 0.4), \\ 10\log_{10} N_N(f) = -15 + 20\log_{10} f, \end{cases} \quad (5)$$

where $N_T$, $N_S$, $N_W$, and $N_N$ are the power spectrum density for turbulence, shipping, waves, and thermal noise, respectively, $s \in [0,1]$ is the shipping factor, $w$ is the wind speed, and the total ambient noise is $10\log_{10}(N_T(f)N_S(f)N_W(f)N_N(f))$.

The PL model of an RF channel can be given by the Maxwell equation [32]

$$L_{i,k}^{\text{RF}} = 8.686\sqrt{\pi\mu\iota f}\, d_{i,k}, \quad (6)$$

where $\mu$ and $\iota$ represent the permeability factor and electrical conductivity of seawater, respectively.

### B. Problem Formulation

Given the problem of establishing an energy-efficient UECN, the main challenge is to simultaneously minimize ERT and EC. Since all system units (USN and AUV) work simultaneously, the practical ERT depends on the longest ERT of system units, which is equivalent to a min-max problem (MMP), whereas the EC stems from all system units. Motivated by this, the original problem (OP) can be defined as the following MOP (OP):

$$\min_{\substack{\mathcal{M},\mathcal{A},\mathcal{B},\mathcal{C},\{\mathbf{u}_j\},\\ \{P_{i,0}\},\{P_{k,i}\},\{P_{l,j}\}}} \max\left\{\{T_i\},\{T_k\},\{\mathbb{S}(|\mathcal{C}|)T_l\},\{\mathbb{S}(|\mathcal{C}|)T_j\}\right\} \quad (7a)$$

$$\min_{\substack{\mathcal{M},\mathcal{A},\mathcal{B},\mathcal{C},\{\mathbf{u}_j\},\\ \{P_{i,0}\},\{P_{k,i}\},\{P_{l,j}\}}} \sum_{i\in\mathcal{A}} E_i + \sum_{k\in\mathcal{B}} E_k + \mathbb{S}(|\mathcal{C}|)\left(\sum_{l\in\mathcal{C}} E_l + \sum_{j\in\mathcal{M}} E_j\right) \quad (7b)$$

s.t. $\mathcal{A}\cup\mathcal{B}\cup\mathcal{C} = \mathcal{N}, \mathcal{A}\cap\mathcal{B}\cap\mathcal{C} = \varnothing,$  (7c)

$D_{\max} \leq z_j^{\text{AUV}}, E_j \leq E_{\max}, \forall j \in \mathcal{M},$  (7d)

$L_{i,0} \in \{L_{i,0}^{\text{UL}}, L_{i,0}^{\text{UA}}, L_{i,0}^{\text{RF}}\}, P_{i,0} \in \{P_{i,0}^{\text{UL}}, P_{i,0}^{\text{UA}}, P_{i,0}^{\text{RF}}\}, \forall i \in \mathcal{A},$  (7e)

$L_{k,i} \in \{L_{k,i}^{\text{UL}}, L_{k,i}^{\text{UA}}, L_{k,i}^{\text{RF}}\}, P_{k,i} \in \{P_{k,i}^{\text{UL}}, P_{k,i}^{\text{UA}}, P_{k,i}^{\text{RF}}\}, \forall k \in \mathcal{B},$  (7f)

$L_{l,j} \in \{L_{l,j}^{\text{UL}}, L_{l,j}^{\text{UA}}, L_{l,j}^{\text{RF}}\}, P_{l,j} \in \{P_{l,j}^{\text{UL}}, P_{l,j}^{\text{UA}}, P_{l,j}^{\text{RF}}\}, \forall l \in \mathcal{C},$  (7g)

$0 \leq P^{\text{UL}} \leq P_{\max}^{\text{UL}}, 0 \leq P^{\text{UA}} \leq P_{\max}^{\text{UA}}, 0 \leq P^{\text{RF}} \leq P_{\max}^{\text{RF}},$  (7h)

where $\mathcal{A}$, $\mathcal{B}$, and $\mathcal{C}$ denote the sets of USNs selecting ERM 1, ERM 2, and ERM 3, respectively, $T_i$, $T_k$, $T_l$, and $T_j$ denote ERT, $E_i$, $E_k$, $E_l$, and $E_j$ denote EC, $D_{\max}$ is the maximum diving depth, $E_{\max}$ is the upper bound of AUV energy, and $\mathbb{S}(|\mathcal{C}|)$ indicates that AUVs are required to assist I-USNs iff $\mathcal{C}$ is not empty.

For simplicity, we consider that $T_i = t_d^{i,0}$, $T_k = t_d^{k,i}$, $T_l = t_d^{l,j}$, and $T_j = t_m^j + \max\{t_d^{l,j}\}, l \in \mathcal{S}_j$, where $\mathcal{S}_j$ is the set of I-USNs served by AUV $j$, $\sum_{j\in\mathcal{M}} |\mathcal{S}_j| = |\mathcal{C}|$, $t_m^j$ is the movement time of AUV $j$, and $t_d^{l,j}$ is given by [33]

$$t_d^{l,j} = \frac{p_{l,j}}{R_{l,j}} + \frac{d_{l,j}}{\omega_{l,j}} + t_g + t_a, \quad (8)$$

where $R_{l,j} = B_{l,j}\log_2\left(1 + \zeta_l P_{l,j} 10^{-(L_{l,j}/10)}(N_0 + I_0)^{-1}\right)$, $B_{l,j}$ is the bandwidth, $P_{l,j}$ is the transmit power, $N_0$ is the noise power, $I_0$ is the variable of interference power, $p_{l,j}$ is the packet size, $\omega_{l,j}$ is the propagation speed (Table I), and $t_g = 0.1\,\text{s}$ and $t_a = 0.1\,\text{s}$ are guard time and preamble time, respectively. Note that $t_d^{i,0}$ and $t_d^{k,i}$ are similar to (8), and $\zeta_l = \left(10\pi |z_l^{\text{USN}}|(1\mu\text{Pa})B_{l,j}\right)^{-1}$ if the UA link is selected; otherwise $\zeta_l = 1$ [12].

Given (8), we let $E_i = P_{i,0}t_d^{i,0}$, $E_k = P_{k,i}t_d^{k,i}$, $E_l = P_{l,j}t_d^{l,j}$, and $E_j = \sum_{l\in\mathcal{S}_j}\left(t_d^{l,j}\zeta_l P_{l,j} 10^{-(L_{l,j}/10)}\right) + E_m^j + \max\{t_d^{l,j}P_{\text{hover}}\}$, where $P_{\text{hover}}$ is the hovering power and $E_m^j$ can be decomposed as

$$E_m^j = e_b^j + e_l^j + e_s^j + e_e^j, \quad (9)$$

where $e_b$, $e_l$, $e_s$, and $e_e$ denote the energy consumed by the buoyancy system, the linear system, the rotational system, and the electronic system, respectively, and $P_{\text{hover}}$ depends on $e_e$.

Specifically, $e_b^j$ is given by [34]

$$e_b^j = \frac{2m_b}{\eta_B\rho}\left(\mathbb{I}\left(\frac{|z_j^{\text{AUV}}|}{2D_{\max}}\right)\rho g D_{\max} + \rho g\mathbb{R}\left(\frac{|z_j^{\text{AUV}}|}{2D_{\max}}\right)D_{\max} + \mathbb{R}\left(\frac{|z_j^{\text{AUV}}|}{2D_{\max}}\right)P_0\right), \quad (10)$$

where $\rho$ is the seawater density, $\eta_B$ is the engine efficiency, $m_b$ is the mass of the net buoyancy, $g$ is the gravity acceleration, and $P_0$ is the atmospheric pressure. Then, $e_l^j$ can be defined as

$$e_l^j = \left(2\mathbb{I}\left(\frac{|z_j^{\text{AUV}}|}{2D_{\max}}\right) + 1\right)\frac{m_l a_l^2\left(d_{j,0}\cos(\theta)\right)^4}{\eta_L}, \quad (11)$$

where $m_l$ is the mass of the movable block, $a_l$ and $\eta_L$ are the linear system constant and efficiency, respectively, and $d_{j,0}$ is the distance between AUV $j$ and USV 0.

To realize the rotation of an AUV, we have

$$e_s^j = \frac{1}{2\eta_S} a_s^2 (\psi_1 - \psi_0)^4, \quad (12)$$

where $a_s$ and $\eta_S$ are the rotational system constant and efficiency, respectively, and $\psi_1$ and $\psi_0$ are the final and initial angles of rotation, respectively.

Finally, to maintain the electronic system, $e_e^j$ is defined as

$$e_e^j = \frac{a_e d_{j,0}}{v_j}, \quad (13)$$

where $a_e$ and $v_j$ are the electronic system constant and the velocity of AUV $j$, respectively.

III. JOINT MODE SELECTION AND AUV DEPLOYMENT

In this section, we first select the optimal ERM for each USN. Second, according to the distribution of I-USNs, the locations and controls of AUVs are jointly optimized to shorten the time for data collection and underwater movement. Finally, the best tradeoff between ERT and EC is obtained using multi-objective optimization.

A. Optimal Mode Selection

To effectively collect emergency data, we need to determine the optimal ERM for a USN and make sure that all USNs' data can be finally forwarded toward the USV, which significantly shortens the time for data transmission and improves the EE of the UECN. Given the locations of geographically fixed USNs, we propose the following stepwise algorithm to determine the optimal mode selection hierarchically.

1) Relay detection: Given the original distribution of all USNs, we determine the USNs in ERM 1 that can work as potential RNs to transfer data.
2) Relay selection: Given the distribution of RNs, we propose an RL-based scheme to determine RNs for USNs in ERM 2 and identify I-USNs.

*1) Optimal Relay Detection*: Obviously, we can assert from (OP) that our goal is to maximize the number of USNs in ERM 1. That is, the most ideal solution to (OP) is that all USNs can transmit data to the USV directly, subject to a given set of SINR constraints. Therefore, given the minimum SINR threshold $\gamma$, (OP) is equivalent to following optimization problem

$$\text{(P1): } \max_{\chi_i} \sum_{i \in \mathcal{N}} \chi_i \quad (14a)$$

$$\text{s.t. } 7(e), 7(h), \chi_i \in \{0,1\}, \gamma \leq \chi_i \left( \frac{\zeta_i P_{i,0} 10^{-(L_{i,0}/10)}}{N_0 + I_0} \right), \forall i \in \mathcal{N}, \quad (14b)$$

where $\chi_i$ denotes a binary variable that is equal to 1 if USN $i$ can connect with the USV directly; otherwise it is set to 0.

Note that the problem above (P1) is non-convex because it is the product of binary variable $\chi_i$ and $\text{SINR}_i$. Even if $\chi_i$ is relaxed to take a value belonging to $\{0,1\}$, the relaxed version of (P1) is still non-convex due to two variables $P_{i,0}$ and $L_{i,0}$ hidden in $\text{SINR}_i$. Thus, this problem (P1) is equivalent to a mixed-integer non-linear problem (MINLP), which is generally NP-hard, and cannot be solved directly.

To make (P1) solvable, we propose a GS-based scheme to solve (P1), as shown in Algorithm 1, which gives the following advantages: 1) The total number of iterations can be reduced to $\sum_{i=0}^{N-1}(N-i)$; 2) The GS can guarantee a near-optimal solution to (P1); 3) The feasible solutions to $P_{i,0}$ and $L_{i,0}$ can be easily obtained by judging whether (14b) is satisfied. Therefore, each USN greedily selects a UCL to maximize its CC, which yields the optimal $L_{i,0}$

$$L_{i,0}^* = \arg\max_{L_{i,0}} \left\{ R_{i,0}^{\text{UL}}, R_{i,0}^{\text{UA}}, R_{i,0}^{\text{RF}} \right\}. \quad (15)$$

Similarly, the optimal $P_{i,0}$ is expressed as

$$P_{i,0}^* = \arg\max_{P_{i,0}} \left\{ R_{i,0}^{\text{UL}}, R_{i,0}^{\text{UA}}, R_{i,0}^{\text{RF}} \right\}, \quad (16)$$

where $P_{i,0}^*$ is given by the following proposition.

*Proposition 1:* The optimal transmit power of USN $i$ is given by $P_{\min} 10^{(L_{i,0}^*/10)}$, where $P_{\min} = \gamma \zeta_i^{-1}(N_0 + I_0)$, $\gamma$ is the minimum SINR threshold, $N_0$ is the noise power, and $I_0$ is the interference power.

*Proof:* Following (14b) and (15)-(16), the optimal $R_{i,0}$ can be obtained when USN $i$ transmits its data to the USV and the remaining USNs stop data transmission, i.e., the optimal SINR is $\zeta_i P_{i,0} \left(10^{(L_{i,0}^*/10)} N_0\right)^{-1}$. However, considering the benefit of USN $k$ ($k \neq i$), USN $i$ cannot transmit data with its maximum transmit power. That is, we need to determine the best tradeoff between the gain ($\text{SINR}_i$) and the interference power to other nodes ($I_i$), which is equivalent to simultaneously maximizing $\text{SINR}_i$ and minimizing $I_i$. Given the minimum received power $P_{\min} = \gamma \zeta_i^{-1}(N_0 + I_0)$, the optimal transmit power of USN $i$ can be determined by comparing the relationship between $\text{SINR}_i^{-1}$ and $I_i$. As shown in Fig. 2, minimizing $\text{SINR}_i^{-1}$ contradicts minimizing $I_i$. Therefore, to obtain the best tradeoff point, $I_i$ decreases until $P_{\min} 10^{(L_{i,0}^*/10)} \leq I_i = P_{i,0}$, which clearly proves the proposition. Furthermore, given the lower bound of $P_{i,0}$, $E_i$ is minimized accordingly.

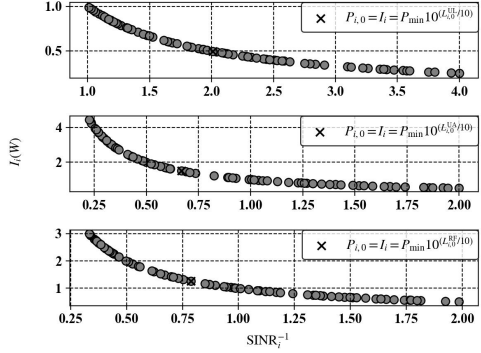

Fig. 2. Interference power versus the reciprocal of an SINR.

Using Proposition 1, (16) can be rewritten as

$$P_{i,0}^* = \arg\max_{P_{i,0}} \left\{ R_{i,0}^* \mid P_{\min} 10^{(L_{i,0}^*/10)} \right\}. \quad (17)$$

In addition, given the fact that USN $i$ has a certain possibility to disconnect from the USV, the outage probability of USN $i$ is introduced into (P1) as a constraint [33]

$$\mathbb{P}_{\text{out}}^{i,0} = 1 - \mathbb{Q}\left( \frac{P_{\min} - 10\log_{10}\left(\zeta_i P_{i,0}^*\right) + L_{i,0}^*}{\sigma} \right), \quad (18)$$

where $\mathbb{P}_{\text{out}}^{i,0} \in \left\{ \mathbb{P}_{i,0}^{\text{UL}}, \mathbb{P}_{i,0}^{\text{UA}}, \mathbb{P}_{i,0}^{\text{RF}} \right\}$ and $\sigma$ is the standard deviation.

To guarantee the continuity and quality of data transmission, given (14b), (18), and the threshold of $\mathbb{P}_{\text{out}}^{i,0}$, denoted by $\varepsilon$, we can relax (P1) as follows:

$$\max_{\mathcal{A}} \sum_{i \in \mathcal{A}} \left( \mathbb{S}\left(\varepsilon - \mathbb{P}_{\text{out}}^{i,0}\right) \mathbb{S}\left( \frac{\zeta_i P_{i,0}^* 10^{-(L_{i,0}^*/10)}}{N_0 + I_0} - \gamma \right) \right) \quad (19a)$$

$$\text{s.t.} \quad I_0 = \alpha_i^{\text{UL}} \sum_{a \in \mathcal{J}} \left( \mathbb{S}\left(\varepsilon - \mathbb{P}_{a,0}^{\text{UL}}\right) \zeta_a P_{a,0}^{\text{UL}} 10^{-(L_{a,0}^{\text{UL}}/10)} \right)$$
$$+ \alpha_i^{\text{UA}} \sum_{b \in \mathcal{K}} \left( \mathbb{S}\left(\varepsilon - \mathbb{P}_{b,0}^{\text{UA}}\right) \zeta_b P_{b,0}^{\text{UA}} 10^{-(L_{b,0}^{\text{UA}}/10)} \right) \quad (19b)$$
$$+ \alpha_i^{\text{RF}} \sum_{c \in \mathcal{L}} \left( \mathbb{S}\left(\varepsilon - \mathbb{P}_{c,0}^{\text{RF}}\right) \zeta_c P_{c,0}^{\text{RF}} 10^{-(L_{c,0}^{\text{RF}}/10)} \right),$$

$$\alpha_i^{\text{UL}}, \alpha_i^{\text{UA}}, \alpha_i^{\text{RF}} \in \{0,1\}, \alpha_i^{\text{UL}} + \alpha_i^{\text{UA}} + \alpha_i^{\text{RF}} = 1, \forall i \in \mathcal{N}, \quad (19c)$$

$$(14b)-(18), \mathcal{J} \cup \mathcal{K} \cup \mathcal{L} = \mathcal{A} \setminus i, \mathcal{J} \cap \mathcal{K} \cap \mathcal{L} = \varnothing, \quad (19d)$$

where $\mathcal{J}$, $\mathcal{K}$, and $\mathcal{L}$ denote the sets of USNs selecting UL, UA, and RF links, respectively, and (19c) ensures that just one type of channel is selected by USN $i$. Given the relaxed (P1), the set of USNs in ERM 1 can be obtained by traversing all USNs in $\mathcal{N}$, as shown in Algorithm 1.

---

**Algorithm 1** Algorithm for Relay Detection

1: **Input:** Set of USNs $\mathcal{N}$
2: **Output:** Set of USNs in ERM 1
3: **Initialize:** $\mathcal{A} \leftarrow \varnothing$
4: **for** $i \in \mathcal{N}$ **do**
5:     Calculate $L_{i,0}^*$, $P_{i,0}^*$, and $\mathbb{P}_{\text{out}}^{i,0}$ using (15), (16), and (18)
6:     Update $\alpha_i^{\text{UL}}$, $\alpha_i^{\text{UA}}$, and $\alpha_i^{\text{RF}}$ subject to (19c)
7:     **for** $j \in \mathcal{N} \setminus i$ **do**
8:         $R_{j,0}^* \leftarrow \max \left\{ R_{j,0}^{\text{UL}}, R_{j,0}^{\text{UA}}, R_{j,0}^{\text{RF}} \right\}$ and select the optimal UCL for USN $j$
9:     **end for**
10:     **if** $\mathbb{S}\left(\varepsilon - \mathbb{P}_{\text{out}}^{i,0}\right) \mathbb{S}\left( \frac{\zeta_i P_{i,0}^* 10^{-(L_{i,0}^*/10)}}{N_0 + I_0} - \gamma \right) = 1$ **then**
11:         Add $i$ to $\mathcal{A}$
12:     **end if**
13: **end for**
14: return $\mathcal{A}$

---

In Algorithm 1, we first determine the optimal link for USN $i \in \mathcal{N}$ using Proposition 1 (steps 5-6). Second, the variable of interference power is obtained using (19b) (steps 7-9). Finally, a potential RN can be identified by judging the value of (19a) (steps 10-12). Since the main loop of Algorithm 1 depends on steps 4-13, the computational complexity can be denoted by $\mathcal{O}(|\mathcal{N}||\mathcal{N} \setminus i|)$, and $|\mathcal{N}|$ controls the convergence rate.

*2) Optimal Relay Selection*: Given the distribution of RNs, we further divide $\mathcal{N} \setminus \mathcal{A}$ into $\mathcal{B}$ and $\mathcal{C}$, where a USN in $\mathcal{B}$ selects a USN in $\mathcal{A}$ as its RN. Following [35], it is proved that the ERT and EC of an AUV are much greater than those of a USN. Thus, we need to maximize the number of USNs in $\mathcal{B}$, by which $T_j$ and $E_j$ can be minimized accordingly. Given (19b)-(19d), our optimization problem can be expressed as follows:

(P2): $$\max_{\mathcal{B},\{P_{k,i}\}} \sum_{k \in \mathcal{N} \setminus \mathcal{A}} \sum_{i \in \mathcal{A}} \left( \beta_i \mathbb{S}\left(\varepsilon - \mathbb{P}_{\text{out}}^{k,i}\right) \mathbb{S}\left(R_{k,i} - \overline{R}\right) \right) \quad (20a)$$

$$\text{s.t.} \quad \beta_i \in \{0,1\}, \sum_{i \in \mathcal{A}} \beta_i = 1, \mathbb{C}(i) \leq C_{\max}, \forall i \in \mathcal{A}, \quad (20b)$$

where $\overline{R} = |\mathcal{N} \setminus \mathcal{A}|^{-1}\left(R_{k,0} + ... + R_{k,|\mathcal{N} \setminus \mathcal{A}|-1}\right)$, $\sum_{i \in \mathcal{A}} \beta_i = 1$ means that USN $k$ only communicates with one RN, $C_{\max}$ is the maximum capacity threshold that equals the number of sub-channels, and $\mathbb{C}(i) \leq C_{\max}$ ensures that the number of USNs connected with RN $i$ cannot exceed $C_{\max}$.

Since (P2) is a dynamic programming problem (DPP), it is challenging to solve (P2) using traditional algorithms, such as ant colony optimization (ACO). Thus, we aim to create a solver to maximize the number of USNs in ERM 2, maximize CC, and minimize the outage probability between USN $k$ and RN $i$. Fortunately, since RL, such as Q-learning, state-action-reward-state-action, and deep Q-network (DQN), heuristically explores solutions from a system state space and continuously updates policies using environmental feedback (rewards), RL-assisted algorithms can provide near-optimal solutions to (P2), yielding a roadmap to optimal or suboptimal solutions [35].

Before determining the optimal RN for a USN, the system state, action, and reward are first defined as follows.

1) State: The set of states at time slot $t$ is defined as

$$\mathbf{s}_t = \left[\mathbb{P}_{k,i}^{\text{UL}}, \mathbb{P}_{k,i}^{\text{UA}}, \mathbb{P}_{k,i}^{\text{RF}}, R_{k,i}^{\text{UL}}, R_{k,i}^{\text{UA}}, R_{k,i}^{\text{RF}}\right], k \in \mathcal{N} \setminus \mathcal{A}, i \in \mathcal{A}, \quad (21)$$

where USN $k$ only selects an RN to maximize its CC and minimize its outage probability.

2) Action: The set of actions at time slot $t$ is expressed as

$$\mathbf{a}_t = [i,...,|\mathcal{A}|], i \in \mathcal{A}, \quad (22)$$

where $\mathbf{a}_t$ is the set of RNs' indexes whose entry denotes the 3D location of RN $i$, and RN $i$ cannot be repeatedly selected by USNs when $\mathbb{C}(i) > C_{max}$.

3) Reward: The reward at each time slot $t$ is given by

$$r_t = \max\left\{\mathbb{S}(\varepsilon - \mathbb{P}_{k,i}^{UL})R_{k,i}^{UL}, \mathbb{S}(\varepsilon - \mathbb{P}_{k,i}^{UA})R_{k,i}^{UA}, \mathbb{S}(\varepsilon - \mathbb{P}_{k,i}^{RF})R_{k,i}^{RF}\right\}. \quad (23)$$

Then, given (21)-(23), USN $k$ selects an action $a_t \in \mathbf{a}_t$ using the current policy $\pi$ and then updates the policy $\pi$ using the generated reward $r_t$, as shown in Algorithm 2, where $a_t$ and $\mathbf{s}_t$ are updated and recorded in a Q-table

$$\mathbf{Q}(a_t,\mathbf{s}_t) \leftarrow \mathbf{Q}(a_t,\mathbf{s}_t) + \alpha\left(r_t + \beta \max_{a \in \mathbf{a}_{t+1}}\{\mathbf{Q}(a_{t+1},\mathbf{s}_{t+1})\} - \mathbf{Q}(a_t,\mathbf{s}_t)\right)$$
$$= (1-\alpha)\mathbf{Q}(a_t,\mathbf{s}_t) + \alpha\left(r_t + \beta \max_{a \in \mathbf{a}_{t+1}}\{\mathbf{Q}(a_{t+1},\mathbf{s}_{t+1})\}\right), \quad (24)$$

where $\mathbf{Q}(a_t,\mathbf{s}_t)$ denotes the value of the Q-table storing system actions and states, $\alpha$ is the learning rate, $\beta$ is the discount factor, and $\mathbf{s}_{t+1}$ is the next state derived from the current state $\mathbf{s}_t$ via the selected action $a_t$. To avoid the problem of local optimality, the $\ell$-greedy scheme is used to select an action [36], i.e.,

$$\pi \leftarrow \begin{cases} a_t = \arg\max_{a \in \mathbf{a}_t}\{\mathbf{Q}(a_t,\mathbf{s}_t)\}, & \text{if } 1-\ell \geq q, \\ a_t = \text{random}\{\mathbf{a}_t\}, & \text{otherwise,} \end{cases} \quad (25)$$

where $\ell$ is a relatively small probability and $q$ is a threshold.

However, with the increase in the number of USNs and the length of state vector $\mathbf{s}_t$, it is inefficient to update and retrieve the Q-table, and the computational complexity increases dramatically. To address this problem, a DQN-based relay selection algorithm is proposed in Algorithm 3, which trains a convolutional neural network (CNN) to approximate $\mathbf{Q}(a_t,\phi_t,\mathbf{w}_t)$. The architecture of DQN comprises 2 convolutional layers (Convs) and 2 fully connected layers (FCs) [37], where the depths of Conv 1, Conv 2, FC 1, and FC 2 are 20, 40, 180, and $|\mathbf{a}_t|$, respectively, and the main training steps are summarized as follows.

1) Input $\phi_t(\mathbf{s}_{t-W},...,\mathbf{s}_t)$ into Conv 1, where $\phi_t(\cdot)$ denotes the reshaping layer that converts $\{\mathbf{s}_{t-W},...,\mathbf{s}_t\}$ into a two-dimension image with a size of $9 \times 9$.

2) Store the transition, denoted by $\{\phi_t, a_t, r_t, \phi_{t+1}\}$, in memory replay pool $\mathcal{D}$ for generating training samples and labels.

3) Extract $B$ training samples from $\mathcal{D}$ at random and update CNN parameters by gradient descent algorithms, as shown in Fig. 3.

Following (24), the loss function can be defined as

$$\mathbb{L}(\mathbf{w}_t) = \mathbb{E}\left[\left(r_t + \beta \max_{a \in \mathbf{a}_{t+1}}\{\mathbf{Q}(a_{t+1},\phi_{t+1},\mathbf{w}_{t-1})\} - \mathbf{Q}(a_t,\phi_t,\mathbf{w}_t)\right)^2\right], \quad (26)$$

where $\mathbf{w}_t$ is the parameter vector, which can be updated by

$$\nabla_{\mathbf{w}_t}\mathbb{L}(\mathbf{w}_t) = -\mathbb{E}\left[\left(r_t + \beta \max_{a \in \mathbf{a}_{t+1}}\{\mathbf{Q}(a_{t+1},\phi_{t+1},\mathbf{w}_{t-1})\} - \mathbf{Q}(a_t,\phi_t,\mathbf{w}_t)\right) \times \nabla_{\mathbf{w}_t}\mathbf{Q}(a_t,\phi_t,\mathbf{w}_t)\right]. \quad (27)$$

In Algorithms 2, we introduce one constraint $\mathbb{C}(i) \leq C_{max}$ to reduce interference power from sub-channels, which changes the length of action set $\mathbf{a}_t$ during the training process (steps 6-8), so the computational complexity of Algorithms 2 relies on $\mathcal{O}(LT|\mathbf{a}_t|)$. By contrast, Algorithm 3 trains a CNN to evaluate $\mathbf{Q}(a_t,\phi_t,\mathbf{w}_t)$, which addresses the issue of high dimensionality and reduces the computational complexity to $\mathcal{O}(LTB)$. Further, integrating Algorithms 1-3 realizes the optimal mode selection. As shown in Algorithm 4, it begins with Algorithm 1, which yields a set of potential RNs ($\mathcal{A}$). Second, the optimal RN of a USN in $|\mathcal{N}/\mathcal{A}|$ can be given by Algorithms 2-3. Based on the above, the computational complexity of Algorithm 4 can be denoted by $\mathcal{O}(|\mathcal{N}||\mathcal{N}\setminus i| + |\mathcal{N}\setminus\mathcal{A}|LTB)$, suggesting that the convergence rate of Algorithm 4 depends on $|\mathcal{N}|$, $\alpha$, and $|\mathbf{a}_t|$.

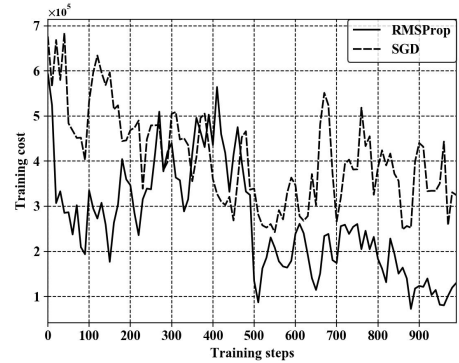

Fig. 3. Comparisons of two popular optimizers, including root mean square propagation (RMSProp) and stochastic gradient descent (SGD).

---

**Algorithm 2** Algorithm for Q-learning-based Relay Selection

1: **Input:** Set of RNs $\mathcal{A}$
2: **Output:** Optimal RN of USN $k$, $\forall k \in \mathcal{N}\setminus\mathcal{A}$
3: **Initialize:** Number of episodes $\leftarrow L$, number of epochs $\leftarrow T$, and $\mathbb{C}(i) \leftarrow 0, \forall i \in \mathcal{A}$
4: **for** $l = 1,...,L$ **do**
5:    **for** $t = 1,...,T$ **do**
6:       **if** $\mathbb{C}(i) > C_{max}$ **then**
7:          Remove $i$ from $\mathbf{a}_t$
8:       **end if**
9:       Select an action $a_t$ using (25)

10:     $\mathbf{s}_t \leftarrow \left[ \mathbb{P}_{k,i}^{\text{UL}}, \mathbb{P}_{k,i}^{\text{UA}}, \mathbb{P}_{k,i}^{\text{RF}}, R_{k,i}^{\text{UL}}, R_{k,i}^{\text{UA}}, R_{k,i}^{\text{RF}} \right]$
11:     Calculate $\mathbf{s}_{t+1}$ and $r_t$ using (21) and (23)
12:     Update $\mathbf{Q}(a_t, \mathbf{s}_t)$ using (24)
13:     $\mathbf{s}_t \leftarrow \mathbf{s}_{t+1}$ and $\mathbb{C}(i) \leftarrow \mathbb{C}(i) + 1$
14:   end for
15: end for
16: return RN $i$

---

**Algorithm 3** Algorithm for DQN-based Relay Selection

1: **Input:** Set of RNs $\mathcal{A}$ and number of state vectors $W$
2: **Output:** Optimal RN of USN $k$, $\forall k \in \mathcal{N} \setminus \mathcal{A}$
3: **Initialize:** Number of episodes $\leftarrow L$, number of epochs $\leftarrow T$, number of training samples $\leftarrow B$, $\mathbb{C}(i) \leftarrow 0, \forall i \in \mathcal{A}$, and $\mathcal{D} \leftarrow \varnothing$
4: **for** $l = 1, ..., L$ **do**
5:   **for** $t = 1, ..., T$ **do**
6:     **if** $\mathbb{C}(i) > C_{\max}$ **then**
7:       Remove $i$ from $\mathbf{a}_t$
8:     **end if**
9:     **if** $t \leq W$ **then**
10:       Select an action $a_t$ at random
11:     **else**
12:       Use $\phi_t$ to yield $\mathbf{Q}(a_t, \phi_t, \mathbf{w}_t)$
13:       Select an action $a_t$ using (25)
14:     **end if**
15:     $\mathbf{s}_t \leftarrow \left[ \mathbb{P}_{k,i}^{\text{UL}}, \mathbb{P}_{k,i}^{\text{UA}}, \mathbb{P}_{k,i}^{\text{RF}}, R_{k,i}^{\text{UL}}, R_{k,i}^{\text{UA}}, R_{k,i}^{\text{RF}} \right]$
16:     Calculate $\mathbf{s}_{t+1}$ and $r_t$ using (21) and (23)
17:     $\mathcal{D} \leftarrow \mathcal{D} \cup \{\phi_t, a_t, r_t, \phi_{t+1}\}$
18:     **for** $b = 1, ..., B$ **do**
19:       Sample $\{\phi_b, a_b, r_b, \phi_{b+1}\}$ from $\mathcal{D}$ at random
20:       Label $y_b \leftarrow r_b + \beta \max_{a \in \mathbf{a}_{b+1}} \{\mathbf{Q}(a_{b+1}, \phi_{b+1}, \mathbf{w}_b)\}$
21:     **end for**
22:     Update $\mathbf{w}_t$ using (27) and $\mathbb{C}(i) \leftarrow \mathbb{C}(i) + 1$
23:   **end for**
24: **end for**
25: return RN $i$

---

**Algorithm 4** Algorithm for Optimal Mode Selection

1: **Input:** Set of USNs $\mathcal{N}$
2: **Output:** Sets of USNs in ERM 1, ERM 2, and ERM 3
3: **Initialize:** $\mathcal{A} \leftarrow \varnothing$, $\mathcal{B} \leftarrow \varnothing$, and $\mathcal{C} \leftarrow \varnothing$
4: Generate $\mathcal{A}$ using **Algorithm 1**
5: **for** $k \in \mathcal{N} \setminus \mathcal{A}$ **do**
6:   Select the RN of USN $k$ using **Algorithms 2-3** and yield the index of RN $i$ by $\arg\max_i \{R_{k,i}\}$
7:   **if** $\mathbb{S}(\varepsilon - \mathbb{P}_{\text{out}}^{k,i}) \mathbb{S}(R_{k,i} - \overline{R}) = 1$ **then**
8:     Add $k$ to $\mathcal{B}$
9:   **end if**
10: **end for**
11: $\mathcal{C} \leftarrow \mathcal{N} \setminus \mathcal{A} \setminus \mathcal{B}$
12: return $\mathcal{A}$, $\mathcal{B}$, and $\mathcal{C}$

---

### B. Optimal AUV Deployment

Using Algorithm 4, I-USNs located far away from the USV are identified. To effectively collect data returned from I-USNs, we jointly optimize the locations and controls of AUVs, which simultaneously minimizes ERT and EC. In addition, as shown in (8)-(13), it is observed that the EC of AUV $j$ shows a linear relationship w.r.t. $t_d$ and $t_m$, meaning that minimizing $t_d$ and $t_m$ yields the minimum EC.

*1) AUV Location Optimization:* The problem of optimizing $t_d^{l,j}$ can be defined as

$$(\text{P3}): \quad \min_{\mathbf{u}_j} \quad t_d^{l,j} = \frac{p_{l,j}}{R_{l,j}} + \frac{d_{l,j}}{\omega_{l,j}} + t_g + t_a, \tag{28}$$

where $d_{l,j} = \sqrt{\left(x_j^{\text{AUV}} - x_l^{\text{USN}}\right)^2 + \left(y_j^{\text{AUV}} - y_l^{\text{USN}}\right)^2 + h_{l,j}^2}$ is the distance between USN $l$ and AUV $j$ and $h_{l,j} = z_j^{\text{AUV}} - z_l^{\text{USN}}$.

To optimally solve (P3), this study proposes the following proposition to jointly maximize $R_{l,j}$ and minimize $d_{l,j}$.

*Proposition 2:* Optimizing (P3) is equivalent to minimizing $d_{l,j}$.

*Proof:* Given (8), (17), and (19b)-(19d), we first redefine (P3) as follows:

$$\max_{d_{l,j}} \quad B_{l,j} \log_2 \left( 1 + \frac{\zeta_l P_{l,j} 10^{-(L_{l,j}/10)}}{N_0 + I_0} \right) + \frac{1}{d_{l,j}} \tag{29a}$$

$$\text{s.t.} \quad (8), (17), (19b) - (19d), \forall l \in \mathcal{S}_j, \forall j \in \mathcal{M} \tag{29b}$$

Next, we regard $d_{l,j}$ as an independent variable in $L_{l,j}$ and derive the first derivatives of $L_{l,j}^{\text{UL}}$, $L_{l,j}^{\text{UA}}$, and $L_{l,j}^{\text{RF}}$, respectively

$$\frac{\partial L_{l,j}^{\text{UL}}}{\partial d_{l,j}} = \left[ \frac{20\pi d_{l,j}^2 \cos(\theta_{l,j})(1 - \cos(\theta_0))}{\ln 10 \eta_T \eta_R \exp(-c(\lambda) d_{l,j}) A_{\text{Rec}}} \right]$$

$$\times \frac{\partial}{\partial d_{l,j}} \left[ \frac{\eta_T \eta_R \exp(-c(\lambda) d_{l,j}) A_{\text{Rec}}}{2\pi d_{l,j}^2 \cos(\theta_{l,j})(1 - \cos(\theta_0))} \right]$$

$$= \left[ \frac{10 d_{l,j}^2 (s_{l,j}/d_{l,j})}{\ln 10 \exp(-c(\lambda) d_{l,j})} \right] \times \frac{\partial}{\partial d_{l,j}} \left[ \frac{\exp(-c(\lambda) d_{l,j})}{d_{l,j}^2 (s_{l,j}/d_{l,j})} \right]$$

$$= -\frac{10(c(\lambda) d_{l,j} + 1)}{\ln 10 d_{l,j}}, \tag{30}$$

$$\frac{\partial L_{l,j}^{\text{UA}}}{\partial d_{l,j}} = \frac{\kappa 10}{\ln 10 d_{l,j}} + \frac{10 \log_{10} \phi(f)}{1000}, \tag{31}$$

$$\frac{\partial L_{l,j}^{\text{RF}}}{\partial d_{l,j}} = 8.686 \sqrt{\pi \mu \iota f}, \tag{32}$$

where $s_{l,j}$ is the horizontal distance between USN $l$ and AUV $j$.

Since (31) and (32) are both greater than 0, we can optimize $L_{l,j}^{UA}$ and $L_{l,j}^{RF}$ by minimizing $d_{l,j}$. However, following (30), $L_{l,j}^{UL}$ decreases monotonously w.r.t. $d_{l,j}$. To optimize $L_{l,j}^{UL}$, we introduce the practical constraint on the communication range, i.e., $d_{l,j} \in [d_{\min}, d_{\max}]$, which yields the lower and upper bounds of $L_{l,j}^{UL}$

$$\begin{cases} L_{\min}^{UL} = 10\log_{10}\left[\dfrac{\eta_T \eta_R \exp(-c(\lambda)d_{\max})A_{\text{Rec}}}{2\pi d_{\max}^2 \cos(\theta_{l,j})(1-\cos(\theta_0))}\right], \\ L_{\max}^{UL} = 10\log_{10}\left[\dfrac{\eta_T \eta_R \exp(-c(\lambda)d_{\min})A_{\text{Rec}}}{2\pi d_{\min}^2 \cos(\theta_{l,j})(1-\cos(\theta_0))}\right], \end{cases} \quad (33)$$

where $\cos(\theta_{l,j}) = s_{l,j}d_{l,j}^{-1}, \forall \theta_{l,j} \in [-\theta_0, \theta_0]$.

Given (33), $L_{l,j}^{UL}$ can be approximated by polynomial curve fitting over $\theta_{l,j}$, subject to $\cos(\theta_0) \leq \cos(\theta_{l,j}) \leq \cos(0°) = 1$, which converts the original curve of $L_{l,j}^{UL}$ into a linear function w.r.t. $\theta_{l,j}$. As shown in Fig. 4, the three-order fitting gives the best performance, indicating that minimizing $h_{l,j}$ yields the minimum $L_{l,j}^{UL}$. Therefore, (29a) is equivalent to the following optimization problem

$$\begin{cases} \max_{\mathbf{u}_j} L_{l,j}^{-1}(d_{l,j}) + d_{l,j}^{-1}, & \text{if a UA or RF link is selected,} \\ \max_{\mathbf{u}_j} L_{l,j}^{-1}(h_{l,j}) + d_{l,j}^{-1}, & \text{if a UL link is selected,} \end{cases} \quad (34)$$

subject to $d_{l,j} \in [d_{\min}, d_{\max}]$ and $\cos(\theta_0) \leq \cos(\theta_{l,j}) \leq 1$, which clearly proves that $R_{l,j}^*$ can be obtained by minimizing $d_{l,j}$.

Following Proposition 2, the optimal location of AUV $j$ is obtained. Further, to effectively solve (P3), we use clustering algorithms to solve the minimum distance between an I-USN and an AUV [38], such as K-means clustering (KMC) [39].

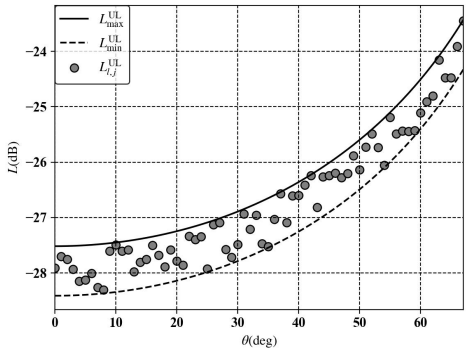

(a) Lower and upper bounds.

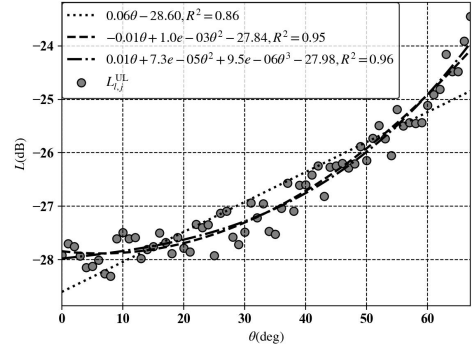

(b) Polynomial curve fitting.

Fig. 4. Approximation of $L_{l,j}^{UL}$.

*2) AUV Control Optimization:* The problem of optimizing $t_m^j$ is equivalent to

$$(\text{P4}): \quad \min_{\mathbf{u}_j, v_j} t_m^j = \frac{d_{j,0}}{v_j}, \quad (35)$$

where $d_{j,0} = \sqrt{(x_j^{AUV} - x_0^{USV})^2 + (y_j^{AUV} - y_0^{USV})^2 + (z_j^{AUV})^2}$ is the distance between USN $j$ and USV 0 and $v_j$ is the velocity of AUV $j$.

To effectively reduce the time for underwater movement, we jointly optimize $d_{j,0}$ and $v_j$, subject to energy constraints (9)-(13). First, the optimization problem w.r.t. $d_{j,0}$ is defined as

$$\min_{\mathbf{u}_j} \sum_{l \in \mathcal{S}_j} d_{l,j} + d_{j,0}. \quad (36)$$

Similar to (28), (36) can be also solved by KMC. In addition, to speed up the convergence of KMC, we initialize the locations of centroids (AUVs) as

$$\arg\min_{\{x_j^{AUV}, y_j^{AUV}, z_j^{AUV}\}} \sum_{l \in \mathcal{S}_j} d_{l,j}. \quad (37)$$

Second, $v_j$ can be optimized by the following proposition.

*Proposition 3:* The optimal $v_j$ is given by

$$v_j^* = \min\{a_e d_{j,0} A_j^{-1}, v_{\max}\}, \quad (38)$$

where $A_j = E_{\max} - e_r^j - e_b^j - e_l^j - e_s^j$, $e_r^j = \sum_{l \in \mathcal{S}_j}\left(t_d^{l,j} \zeta_l P_{l,j} 10^{-(L_{l,j}/10)}\right)$, and $v_{\max}$ is the upper bound of velocity.

*Proof:* From the perspective of the energy-efficient control, we need to minimize $t_m$ subject to a given set of velocity and energy constraints, so the optimization problem is defined as follows:

$$\min_{v_j} -v_j \quad (39a)$$

$$\text{s.t.} \quad e_r^j + E_m^j \leq E_{\max}, 0 \leq v_j \leq v_{\max}, \forall j \in \mathcal{M} \quad (39b)$$

Obviously, since (39a) is a convex optimization problem, its Lagrangian function can be expressed as

$$\mathcal{L}(v_j, \vartheta_1, \vartheta_2, \vartheta_3) = -v_j + \vartheta_1(a_e d_{j,0} v_j^{-1} - A_j) + \vartheta_2(v_j - v_{\max}) + \vartheta_3(-v_j), \quad (40)$$

where $A_j = E_{\max} - e_r^j - e_b^j - e_l^j - e_s^j \geq 0$, $\vartheta_1$, $\vartheta_2$, and $\vartheta_3$ denote

three Lagrange's multipliers, and the Karush-Kuhn-Tucker (KKT) conditions satisfy

$$\nabla_{v_j}\mathcal{L}(v_j,\vartheta_1,\vartheta_2,\vartheta_3) = -1-\vartheta_1 a_e d_{j,0}v_j^{-2}+\vartheta_2-\vartheta_3=0, \quad (41)$$

$$\vartheta_1(a_e d_{j,0}v_j^{-1}-A_j)=0, \vartheta_1\geq 0, a_e d_{j,0}v_j^{-1}-A_j\leq 0, \quad (42)$$

$$\vartheta_2(v_j-v_{\max})=0, \vartheta_2\geq 0, v_j-v_{\max}\leq 0, \quad (43)$$

$$\vartheta_3(-v_j)=0, \vartheta_3\geq 0, -v_j\leq 0, \quad (44)$$

where (41) means the necessary condition of a feasible solution and (42)-(44) denote complementary slackness conditions. By solving (41)-(44), we consider the following eight cases, where case 3 and case 7 clearly prove the proposition.

1) Given $\vartheta_1=\vartheta_2=\vartheta_3=0$, (41) does not hold.
2) Given $\vartheta_1=\vartheta_2=0$ and $\vartheta_3\neq 0$, (41) does not hold.
3) Given $\vartheta_1=\vartheta_3=0$ and $\vartheta_2\neq 0$, we have $v_j^*=v_{\max}$, which holds iff $v_{\max}\geq a_e d_{j,0}A_j^{-1}$.
4) Given $\vartheta_2=\vartheta_3=0$ and $\vartheta_1\neq 0$, (41) does not hold.
5) Given $\vartheta_1=0$, $\vartheta_2\neq 0$, and $\vartheta_3\neq 0$, (43) contradicts (44).
6) Given $\vartheta_2=0$, $\vartheta_1\neq 0$, and $\vartheta_3\neq 0$, (42) contradicts (44).
7) Given $\vartheta_3=0$, $\vartheta_1\neq 0$, and $\vartheta_2\neq 0$, we have $v_j^*=v_{\max}$, which holds iff $v_{\max}=a_e d_{j,0}A_j^{-1}$.
8) Given $\vartheta_1\neq 0$, $\vartheta_2\neq 0$, and $\vartheta_3\neq 0$, no $v_j$ simultaneously satisfies the inequalities in (41)-(43).

By solving (P1)-(P4), variables $\mathcal{A}$, $\mathcal{B}$, $\mathcal{C}$, $\{\mathbf{u}_j\}$, $\{P_{i,0}\}$, and $\{P_{k,i}\}$ have been optimized. In the next section, AC-MEA is introduced to optimize $\mathcal{M}$ and $\{P_{l,j}\}$.

### C. Optimal Time-Energy Balancing

Following the basic framework proposed in [40], a modified MEA is proposed to determine the number of AUVs and the transmit power of I-USNs, by which the best tradeoff between ERT and EC is obtained, as shown in Algorithm 5. Specifically, considering (P1)-(P4), we cannot simultaneously minimize the maximum ERT and the total EC, i.e., increasing the number of AUVs shortens ERT but consumes more energy, which yields the contradiction between (7a) and (7b). Motivated by this, we can rewrite (OP) as follows:

$$\min_{x,y} \max\{\{T_i\},\{T_k\},\{\mathbb{S}(|\mathcal{C}|)T_l\},\{\mathbb{S}(|\mathcal{C}|)T_j\}\} \quad (45a)$$

$$\min_{x,y} \sum_{i\in\mathcal{A}}E_i+\sum_{k\in\mathcal{B}}E_k+\mathbb{S}(|\mathcal{C}|)\left(\sum_{l\in\mathcal{C}}E_l+\sum_{j\in\mathcal{M}}E_j\right) \quad (45b)$$

$$\text{s.t.} \quad (7a)-(7h),(8),(19b)-(19d), \quad (45c)$$

$$P_{\min}10^{(L_{l,j}^r/10)}-y_l\leq 0, y_l-P_{\max}^*\leq 0, \forall y_l\in\mathbf{y}, \quad (45d)$$

$$(38),(39b), 0\leq\varepsilon-\mathbb{P}_{\text{out}}^{l,j}, \forall l\in\mathcal{C}, \forall j\in\mathcal{M}, \quad (45e)$$

$$1\leq x\leq|\mathcal{C}|, E_j-E_{\max}\leq 0, \forall j\in\mathcal{M}, \quad (45f)$$

where $x$ denotes the number of AUVs, $\mathbf{y}$ denotes the transmit power of I-USNs, $x\leq|\mathcal{C}|$ indicates that the maximum number of AUVs is $|\mathcal{C}|$, and the number of I-USNs served by AUV $j$ cannot exceed $C_{\max}$.

Using the Tchebycheff's decomposition, (45a)-(45f) can be further simplified as follows:

$$\min_{\mathbf{v}} \mathcal{F}(\mathbf{v}|\boldsymbol{\lambda},\mathbf{z}^*)=\max\{\lambda_i|f_i(\mathbf{v})-z_i^*|\} \quad (46a)$$

$$\text{s.t.} \quad 1\leq i\leq 2, (45c)-(45f), \quad (46b)$$

where $\mathbf{v}$ is $[x,\mathbf{y}]$, $\mathbf{z}^*=[z_1^*,z_2^*]$ is the reference vector that can minimize (45a) and (45b), $z_i^*$ is the optimal value of objective function $i$, $\boldsymbol{\lambda}=[\lambda_1,\lambda_2]$ is the weight vector, and the interval between $\mathcal{F}(\mathbf{v}|\boldsymbol{\lambda},\mathbf{z}^*)$ and $\mathcal{F}(\mathbf{v}|\boldsymbol{\lambda}',\mathbf{z}^*)$ is minimized to yield the following set of non-dominated solutions

$$\mathcal{E}=\left\{\mathcal{F}^*(\mathbf{v}_i)^{-1}\right\}, \quad (47)$$

where $\mathbf{v}_i$ is the solution to objective function $i$ and $\mathcal{F}(\cdot)$ is the function value.

Given the results derived from the PF, we need to determine the best tradeoff point. To this end, based on the basic idea of compromise programming [41], we select EE as a performance indicator to evaluate the optimality of nondominated solutions, which can be expressed as the reciprocal of the weighted sum of normalized ERT and normalized EC, i.e.,

$$\max_{\substack{\{T_j\},\{T_i\},\{T_k\},\{T_l\},\\ \{E_j\},\{E_i\},\{E_k\},\{E_l\}}} \left[\zeta\mathbb{N}\left(\max\{\{T_i\},\{T_k\},\{\mathbb{S}(|\mathcal{C}|)T_l\},\{\mathbb{S}(|\mathcal{C}|)T_j\}\}\right)\right.$$
$$\left.+(1-\zeta)\mathbb{N}\left(\sum_{i\in\mathcal{A}}E_i+\sum_{k\in\mathcal{B}}E_k+\mathbb{S}(|\mathcal{C}|)\left(\sum_{l\in\mathcal{C}}E_l+\sum_{j\in\mathcal{M}}E_j\right)\right)\right]^{-1}, \quad (48)$$

where $\zeta$ is the priority weight that equals 0.5, indicating that saving ERT is as important as reducing EC.

Finally, by solving (48), the optimal number of AUVs and the optimal transmit power of I-USNs can be obtained, which simultaneously minimizes (45a) and (45b), and the EE of the UECN can be improved accordingly.

---

**Algorithm 5** AC-MEA

1: **Input:** MOP (45a)-(45b), number of subproblems $X$, population size $Y$, weight vector $\{\boldsymbol{\lambda}_1,...,\boldsymbol{\lambda}_Y\}$, and number of neighbourhood weight vectors $Z$
2: **Output:** $\mathcal{E}$
3: **Initialize:** $\{i_1,...,i_Z\}$, $i\in\{1,...,Y\}$, $\{\boldsymbol{\lambda}_{i_1},...,\boldsymbol{\lambda}_{i_Z}\}$, $\{\mathbf{v}_1,...,\mathbf{v}_Y\}$, $\{z_1^*,...,z_X^*\}$, $F\leftarrow 1$, and $\mathcal{E}\leftarrow\varnothing$
4: **for** $i=1,...,Y$ **do**
5: $\quad \mathbf{v}'\leftarrow$ Genetic Operators$(\mathbf{v}_m,\mathbf{v}_n)$ $\forall m$ $n\in\{i_1,\ i_Z\}$
6: $\quad \mathbf{v}''\leftarrow$ Repair and Improve$(\mathbf{v}')$
7: $\quad$ **while** $F=1$ **do**
8: $\quad\quad$ Generate $u$ groups $\{\mathcal{P}_1,...,\mathcal{P}_u\}$ using KMC
9: $\quad\quad$ **for** $\mathcal{P}_v\in\{\mathcal{P}_1,...,\mathcal{P}_u\}$ **do**
10: $\quad\quad\quad$ **if** $|\mathcal{P}_v|>C_{\max}$ **then**
11: $\quad\quad\quad\quad$ Update $\mathbf{v}''$, $u\leftarrow u+1$, and $F\leftarrow 1$

```
12:             break
13:         else
14:             F ← 0
15:         end if
16:       end for
17:     end while
18:     for j = 1,..., X do
19:       if z_j^* > f_j(v″) then
20:         z_j^* ← f_j(v″)
21:       end if
22:     end for
23:     for k ∈ {i_1,...,i_Z} do
24:       if F(v″|λ_k, z^*) ≤ F(v_k|λ_k, z^*) then
25:         v_k ← v″ and F(v_k) ← F(v″)
26:       end if
27:     end for
28:     Remove all vectors dominated by F(v″) from E
29:     Add v″ to E if no vectors in E dominate F(v″)
30: end for
31: return E
```

In Algorithm 5, steps 5-6 first generate the initial solution to (OP). Next, an adaptive KMC loop is performed to repair and improve $\mathbf{v}'$, which also speeds up the convergence rate of AC-MEA. Then, variables $\mathbf{z}^*$, $\mathbf{v}_k$, and $\mathcal{F}(\mathbf{v}_k)$ are updated using steps 18-27. Finally, we remove all dominated solutions from $\mathcal{E}$ and add non-dominated solutions to $\mathcal{E}$. Since $\{\mathcal{P}_1,...,\mathcal{P}_u\}$ changes with $u$ (steps 8-17), the computational complexity of Algorithm 5 depends on $\mathcal{O}(XYZ \max\{|\mathcal{P}_v|\})$, and the convergence rate is controlled by the population size.

## IV. SIMULATION RESULTS AND DISCUSSIONS

In our simulations, the number of USNs belongs to {50, 100, 150, 200, 250, 300}, and USNs are randomly distributed within a 1 km × 1 km geographic area. The simulation parameters are presented in Tables II-IV, where the upper bounds of transmit power are from [8] and [42] and the parameters of environments and AUV performance are given by [14] and [34], respectively. Moreover, the used optimizer is RMSProp, where the learning rate is 0.01, the discount rate is 0.9, and the memory-pool size is 2000. To verify the effectiveness of our approach, we compare the following benchmark schemes: 1) Acoustic scheme [14]: use acoustic links to transmit data without deploying AUVs; 2) Optical scheme [43]: a modified Kuhn-Munkres algorithm is used to assign RNs for USNs over optical links without using AUVs; 3) Hybrid scheme [44]: utilize acoustic and optical links for uplink and downlink communications, respectively. By contrast, we first determine the optimal ERM for a USN. Second, the locations and velocities of AUVs are optimized to minimize ERT and EC. Finally, the best time-energy balancing is achieved using AC-MEA. Note that all simulation results in this section are averaged over a large number of independent experiments.

TABLE III: UNDERWATER COMMUNICATION PARAMETERS

| Parameter | Description | Value |
| --- | --- | --- |
| $\eta_T$ | Optical efficiency of a transmitter | 0.9 |
| $\eta_R$ | Optical efficiency of a receiver | 0.9 |
| $c(\lambda)$ | Extinction coefficient | 0.1514 [8] |
| $A_{Rec}$ | Receiver aperture area | 0.01 m$^2$ |
| $\theta_0$ | Laser beam divergence angle | 68.0 ° |
| $N_0$ | Noise power | -130.0 dBm |
| $P_{max}^{UL}$ | Maximum UL transmit power | 1.0 W |
| $P_{max}^{UA}$ | Maximum UA transmit power | 4.5 W |
| $P_{max}^{RF}$ | Maximum RF transmit power | 3.0 W [42] |
| $p$ | Packet size | 1.0 Mb |
| $\varepsilon$ | Outage probability threshold | 0.01 |
| $\gamma$ | SINR threshold | $\mathbb{P}(\varepsilon)^{-1}$ |
| $B$ | Bandwidth | 1 kHz |

TABLE IV: ENVIRONMENTAL PARAMETERS

| Parameter | Description | Value |
| --- | --- | --- |
| $\kappa$ | Spreading coefficient | 1.5 |
| $s$ | Shipping factor | 0.5 |
| $w$ | Wind speed | 0.5 m/s |
| $\mu$ | Permeability factor | 1.256×10$^{-6}$ H/m |
| $\iota$ | Electrical conductivity | 0.01 S/m [14] |
| $\rho$ | Seawater density | 1027.0 kg/m$^3$ |

TABLE V: AUV PERFORMANCE PARAMETERS

| Parameter | Description | Value |
| --- | --- | --- |
| $\eta_B$ | Engine efficiency | 0.7 |
| $m_b$ | Mass of the net buoyancy | 0.494 kg |
| $g$ | Gravity acceleration | 9.8 m/s$^2$ |
| $D_{max}$ | Maximum depth | 200.0 m |
| $P_0$ | Atmospheric pressure | 101.325 kPa |
| $m_l$ | Mass of the movable block | 11.0 kg |
| $a_l$ | Linear system constant | 0.1 |
| $\eta_L$ | Linear system efficiency | 0.85 |
| $a_s$ | Rotational system constant | 1.0 |
| $\eta_S$ | Rotational system efficiency | 0.85 |
| $a_e$ | Electronic system constant | 1.5 |
| $v_{max}$ | Maximum velocity | 1.0 m/s [34] |

Fig. 5 shows the results of optimal mode selection. First, Algorithm 1 increases the percentage of USNs selecting ERM 1 as compared to a hybrid UL-UA network, increasing by 30%, 31%, 40%, 34%, 19%, and 23%, respectively. This indicates the advantage of using multiple channels and the effectiveness of Proposition 1. Second, using RL schemes, the percentage of USNs selecting ERM 2 can be increased by 18%, 7%, 4%, 5%, 12%, and 7%, respectively, decreasing the number of I-USNs served by AUVs. After relay detection and selection, therefore, the number of I-USNs decreases by 24, 36, 66, 76, 78, and 87, respectively. It should be noted that the percentage of I-USNs is equivalent to the loss ratio caused by not using AUVs, meaning that the USV cannot receive data returned from I-USNs without the assistance of AUVs, and Fig. 5(c) illustrates an example of node distribution with 50 nodes, where a USN selects an RN to transfer data to the USV and an I-USN communicates with an AUV.

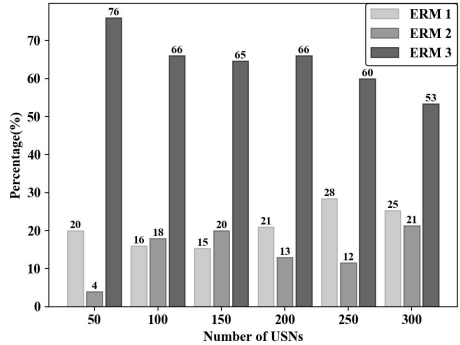

(a) Hybrid scheme.

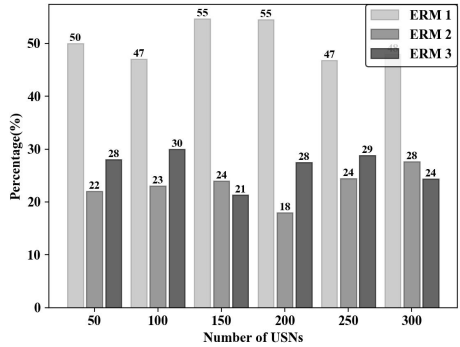

(b) Proposed scheme.

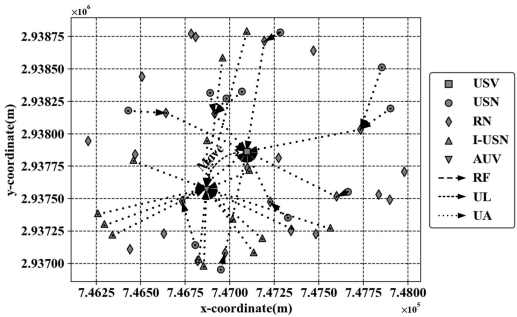

(c) Illustration of node distribution ($N = 50$)

Fig. 5. Optimal mode selection.

Fig. 6 shows the CC averaged over USNs selecting ERM 2. First, Fig. 6(a) plots the convergence rates with different node sizes, indicating that the increase in $N$ also increases the number of potential RNs. Thus, DQN needs to spend more time on relay selection. Meanwhile, since the throughput of RN $i$ is limited, the length of action space **a** changes during the training process of DQN, causing the perturbation of convergence. Second, Fig. 6(b) compares the performance of Q-learning, DQN, particle swarm optimization (PSO), and differential evolution (DE). It is observed that the four algorithms give similar performance when $N$ is relatively small and DQN achieves higher CC with the increase in $N$. Here, the performance difference between RL (e.g., Q-learning, DQN) and evolutionary algorithms (e.g., PSO, DE) depends on two factors: 1) RL follows gradient descent to update the value function, which guarantees the optimality of solutions; 2) RL makes full use of all system states of samples during iteration. Third, in terms of computational cost, the computational time of DQN is longer than other solvers since it needs to update CNN weights, as shown in Table VI. Finally, Fig. 6(c) proves that our proposed algorithm outperforms the other three benchmark schemes, indicating that integrating the merits of multiple links can effectively mitigate the underwater communication latency between a USN and an RN.

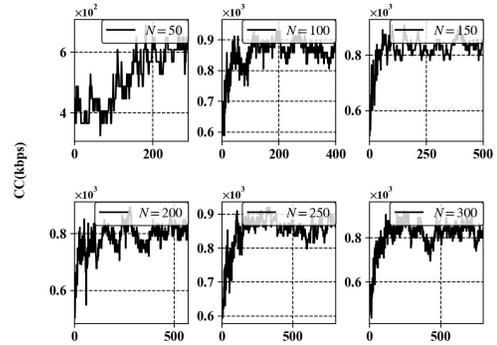

(a) Convergence rates of CC.

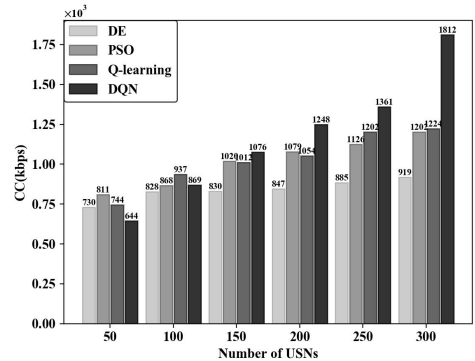

(b) Comparisons of relay selection algorithms.

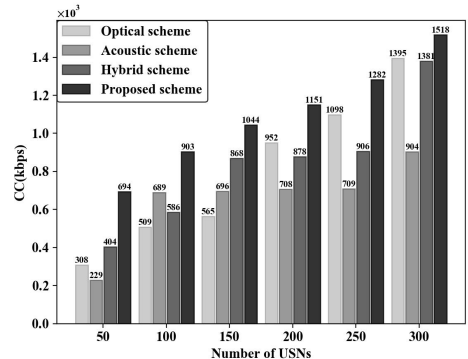

(c) Comparisons of different schemes.

Fig. 6. Optimal relay selection.

TABLE VI: COMPUTATIONAL TIME (S) PER RUN OF DIFFERENT SOLVERS

| $N$ | Q-learning | DQN | PSO | DE |
| --- | --- | --- | --- | --- |
| 50 | 8.01 | 17.65 | 4.83 | 9.03 |
| 100 | 45.50 | 120.51 | 24.70 | 46.22 |
| 150 | 121.09 | 401.30 | 69.52 | 139.01 |
| 200 | 573.71 | 824.36 | 314.95 | 543.63 |
| 250 | 812.80 | 1327.26 | 517.87 | 781.08 |
| 300 | 1095.74 | 1782.65 | 903.67 | 1148.95 |

Fig. 7 shows the results of optimal AUV deployment. First, we compare our deployment scheme with mean-shift clustering (MSC) and density-based spatial clustering of applications with noise (DBSCAN). It is observed from Fig. 7(a) that the optimal numbers of AUVs given by our algorithm in the six scenarios vary from 1 to 2, which suggests that our deployment scheme effectively reduces the number of AUVs by jointly considering the 3D locations of I-USNs and the overall time-energy tradeoff. Second, Fig. 7(b) plots the maximum ERT in different working modes, where the practical ERT relies on the maximum ERT of system units. In addition, Figs. 7(c)-7(d) prove that our proposed deployment scheme simultaneously reduces the maximum ERT and the total EC, decreasing average ERT and average EC by 19.4% and 31.4%, respectively.

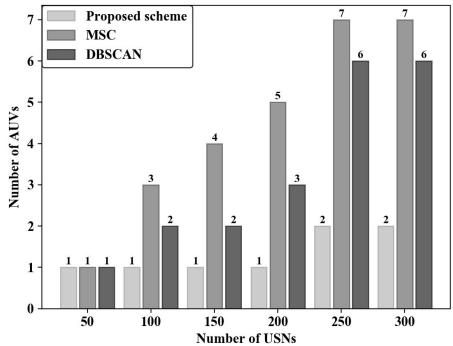
(a) Optimal number of AUVs.

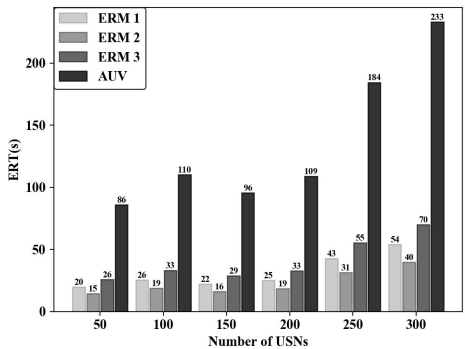
(b) ERT in different modes.

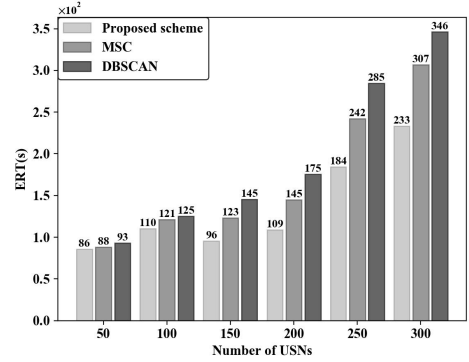
(c) Comparisons of ERT.

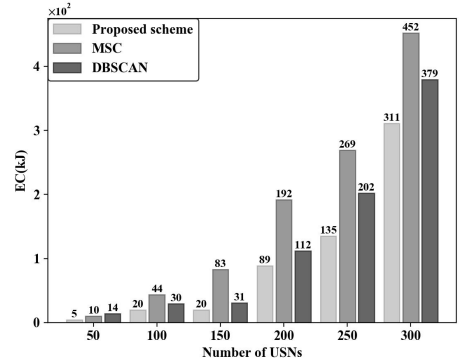
(d) Comparisons of EC.

Fig. 7. Optimal AUV deployment.

Fig. 8 shows the optimal solution to (OP). Since (45a) is an MMP, the optimal ERT equals the maximum ERT of different units, whereas (45b) is a minimization problem, so its optimal solution depends on the cumulative sum of the EC of all system units. To maximize the EE of the UECN, we need to determine the optimal number of AUVs and the optimal transmit power of I-USNs to simultaneously minimize the maximum ERT and the total EC. First, as shown in Fig. 8(a), the PFs consist of many sparse suboptimal points. Next, according to the convergence of PFs, we can find out the best tradeoff point by maximizing EE. As shown in Fig. 8(b), since saving ERT is as crucial as reducing EC, we normalize ERT and EC to be between 0 and 1 and equate the priority weight to 0.5. Then, the corresponding solutions to EE are shown in Fig. 8(c), where the best tradeoff point is obtained when EE achieves its maximum. Finally, to verify the effectiveness of AC-MEA, we compare the following four algorithms: 1) non-dominated sorting genetic algorithm (NSGA-II) 2) NSGA-III 3) PSO and 4) adaptive geometry estimation-based MEA (AGE-MEA). Fig. 8(d) shows that our proposed AC-MEA outperforms the other four solvers in terms of EE, achieving 2.19, 1.97, 1.95, 1.93, 1.88, and 1.85, respectively, and Fig. 8(e) indicates that our proposed approach increases average EE by 0.14, 0.15, and 0.49, respectively.

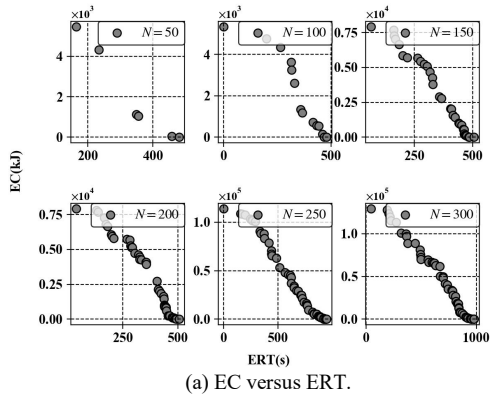

(a) EC versus ERT.

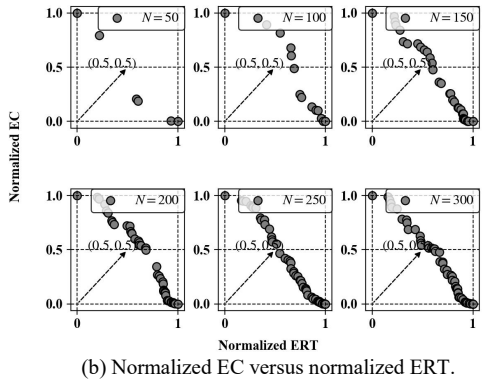

(b) Normalized EC versus normalized ERT.

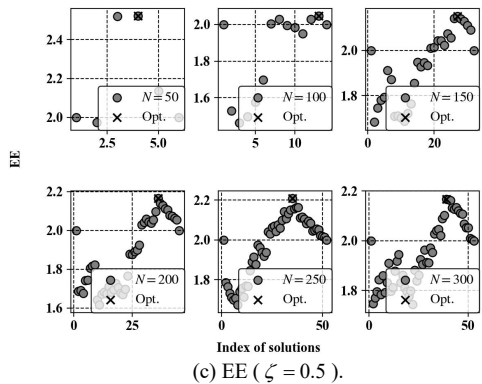

(c) EE ( $\zeta = 0.5$ ).

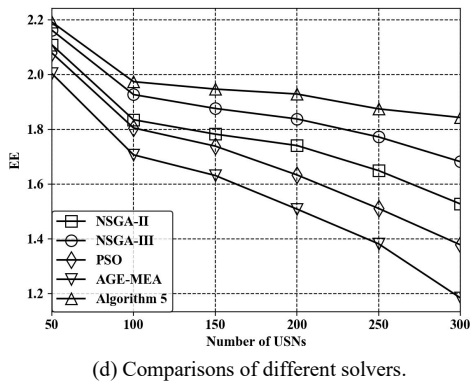

(d) Comparisons of different solvers.

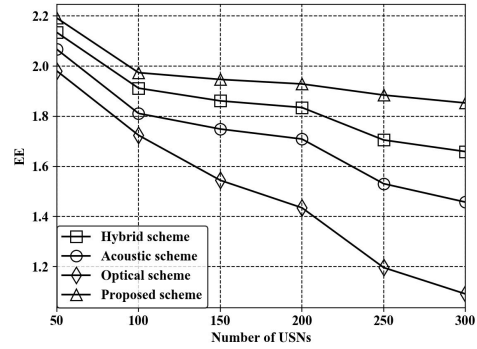

(e) Comparisons of different schemes.

Fig. 8. Optimal solution to (OP).

## V. CONCLUSION

In this study, we built an energy-efficient UECN to collect underwater emergency data. First, the optimal ERM of a USN is determined using relay detection and selection. Specifically, RNs were identified by GS, and the best relay selection policy was given by RL, which strengthened network connectivity and diminished the number of I-USNs served by AUVs. Second, according to the distribution of I-USNs, AUVs were dispatched to assist I-USNs in data transmission, which not only shortened ERT but also reduced EC. Finally, the best tradeoff between the maximum ERT and the total EC was achieved by maximizing EE. Numerical results proved the effectiveness of our proposed approach.

However, there exist potential problems not addressed in this study, some of which are worthy of an in-depth discussion in our future works. Motivated by [45], it is promising to combine intelligent reflecting surfaces (IRSs) with our proposed system, i.e., AUVs are equipped with IRSs to reflect impinging signals with tunable reflection coefficients, including an amplitude and a phase shift, which remarkably enhances the communication throughput without the need for dispatching extra AUVs. Next, to optimally coordinate the reflections of all IRS elements, RL may provide near-optimal solutions without prior knowledge of underwater environments. Furthermore, an IRS has advantages, such as light weight and flexibility, so it becomes feasible to deploy IRSs over the sea surface to support very long-distance communications, leading to discussions about the placement of IRSs and the shortest-path selection.